\documentclass[aps,prl,twocolumn,superscriptaddress,amsmath,amssymb,floatfix,showpacs]{revtex4-1}
\usepackage{enumitem}
\usepackage{graphicx}
\usepackage[caption=false]{subfig}
\usepackage{mathtools}
\def\tr{\mathrm{tr}}

\def\GeV{\mathrm{GeV}}
\def\Complexes{\mathbb{C}}

\def\Higgs{\mathrm{H}}

\def\cV{\mathcal{V}}
\def\tI{t_{\mathrm{I}}}
\def\tW{t_{W}}
\def\tP{t_{\mathrm{grav}}}
\def\tH{t_{\mathrm{H}}}
\def\epsI{\epsilon_{\mathrm{w}}}
\def\SO{\mathrm{SO}}
\def\SU{\mathrm{SU}}
\def\U{\mathrm{U}}
\def\Spin{\mathrm{Spin}}
\def\su{\mathfrak{su}}
\def\eq{\begin{equation}}
\def\en{\end{equation}}
\def\ntimes{\,{\times}\,}
\def\max{\mathrm{max}}
\def\epsilonb{\epsilon_{b}}
\def\epsilona{\epsilon_{a}}

\DeclareMathOperator{\cn}{cn}

\def\EW{\scriptscriptstyle\mathrm{EW}}
\def\zb{z}
\def\yb{y}
\def\kb{k}
\def\Kb{K}
\def\ka{k_{a}}
\def\Ka{K_{a}}

\def\kB{k_{B}}

\def\gtwo{g}
\def\vH{v}
\def\lambdaH{\lambda}

\def\dVol{d^{3}\hat x}
\urldef{\footurla}\url{https://cocalc.com/share/3c1ab84c375769c3460251d4f2bd43461c1211b5/Supplemental_material/}
\urldef{\footurlb}\url{https://www.physics.rutgers.edu/pages/friedan/papers/2020/Origin/Supplemental_material/}
\begin{document}
\title{Origin of cosmological temperature}
\author{Daniel Friedan}
\email[]{dfriedan@gmail.com}
\homepage[]{\path{www.physics.rutgers.edu/~friedan/}}
\affiliation{New High Energy Theory Center and Department of Physics and Astronomy,
Rutgers, The State University of New Jersey,
Piscataway, New Jersey 08854-8019 U.S.A.}
\affiliation{Science Institute, The University of Iceland,
Reykjavik, Iceland
%\thanks{}
}
\date{\today}
\begin{abstract}
%
%0123456789012345678901234567890123456789012345678901234567890123456789
A classical solution of the Standard Model + General
Relativity is given by an elliptic function whose periodicity in
imaginary time is the origin of cosmological temperature.  
Nothing beyond the Standard Model is assumed.
The
solution is a $\Spin(4)$-symmetric universe expanding prior to the
electroweak transition.  
A rapidly oscillating $\SU(2)$ gauge field holds the Higgs field to
$0$ with strength inversely proportional to the scale factor $a$.
When $a$ reaches $a_{\EW}$ the solution becomes unstable and
the electoweak transition begins.  $a_{\EW}$ is the only free
parameter in the solution.  The temperature at $a_{\EW}$ is
$m_{H}/{(6\pi)^{1/2}} = 28.8\,\text{GeV} = 3.34\ntimes 10^{14}\,
\text{K}$ whatever the value of $a_{\EW}$.
\end{abstract}
%\keywords{}
%
\maketitle
Consider the classical equations of motion of the Standard Model 
combined with General Relativity.
Assume (1) the only nonzero fields are
the space-time metric $g_{\mu\nu}(x)$,
the $\SU(2)$ gauge field $B_{\mu}(x) \in \su(2)$,
and the Higgs scalar field $\phi(x)\in \Complexes^{2}$,
(2) space is the 3-sphere $S^{3}$,
and (3) the universe is $\Spin(4)$-symmetric.
Details of the calculations are shown in 
the supplemental material
\footnote{
See the accompanying Supplemental Material consisting of
a note {\it Calculations for "Origin of Cosmological Temperature"}
and a SageMath notebook performing numerical calculations.
The Supplemental Material is also available at \footurla
and at \footurlb.
}.
% https://cocalc.com/share/3c1ab84c375769c3460251d4f2bd43461c1211b5/Supplemental\_material/
% The Supplemental Material is also available at\\ 
% {\tt https://cocalc.com/share/}\\
% {\tt 3c1ab84c375769c3460251d4f2bd43461c1211b5/}\\
% {\tt Supplemental\_material/}\\
% {\tt 
% https://www.physics.rutgers.edu/pages/friedan/}\\
% {\tt papers/2020/Origin/Supplemental\_material/}
% %\href{https://cocalc.com/share/3c1ab84c375769c3460251d4f2bd43461c1211b5/Supplemental_material/}{cocalc.com}
%\cite{cosmonoteIIv3}.

The action (in units of $\hbar$ with  $c=1$) is
\eq
\label{eq:SMGR}
\begin{split}
\frac1\hbar S &=\int
\bigg [
\frac1\hbar \frac{1}{2\kappa} R
+
\frac1{2 \gtwo ^{2}}
\tr( F_{\mu\nu} F^{\mu\nu})
\\
&
{}-D_{\mu}\phi^{\dagger} D^{\mu}\phi
-  \frac12\lambdaH^{2}\left(\phi^{\dagger}\phi 
-  \frac12 \vH ^{2}\right)^{2}
\bigg]\, \sqrt{-g} \,d^{4}x
\end{split}
\en
Space  is the unit 3-sphere  $S^{3}$
identified with $\SU(2)$ by
\eq
\hat x \in S^{3}
\:\:\:\longleftrightarrow\:\:\:
g_{\hat x} = \hat x^{4} \mathbf1 + \hat x^{a}  i^{-1} \sigma_{a} \in\SU(2)
\en
$\sigma_{1,2,3}$ being the Pauli matrices.
Points in space-time are $x=(x^{0},{\hat x}) 
= (x^{0},g_{\hat x})$.
$\Spin(4)=\SU(2)_{L}{\ntimes} \SU(2)_{R}$ acts as $\SO(4)$ on 
space-time by
\eq
x = (x^{0},g_{\hat x})
\mapsto
Ux =  (x^{0},g_{L} g_{\hat x} g_{R}^{-1})
\en
An $\SO(4)$-symmetric metric has the
conformally flat form
\eq
\label{eq:metric}
g_{\mu\nu}dx^{\mu}dx^{\nu} =  a(T)^{2} \left(- dT^{2} + 
ds^{2}_{S^{3}} \right)
\en
$ds^{2}_{S^{3}}$ being the metric of the unit 3-sphere.

As a symmetry ansatz
for solving the equations of motion,
suppose  $\Spin(4)$ acts on $\phi(x)$ and $B_{\mu}(x)$ by
\eq
\begin{aligned}
\phi(x) &\mapsto g_{L}^{-1} \phi(Ux)
\\[1ex]
B_{\mu}(x) dx^{\mu} &\mapsto g_{L}^{-1} B_{\mu}( Ux)\,g_{L}\, d(Ux)^{\mu}
\end{aligned}
\en
The $\Spin(4)$-symmetric scalar and gauge fields are
\eq
\label{eq:invariantBmu}
\phi(x) = 0
\qquad
B_{\mu}(x)dx^{\mu} = \left[1 + b(T)\right] 
\frac12 g_{\hat x} dg_{\hat x}^\dagger 
\en
Cosmology is described by the functions
$a(T)$ and $b(T)$.

\begin{figure}[htp]
\includegraphics[scale=0.5]{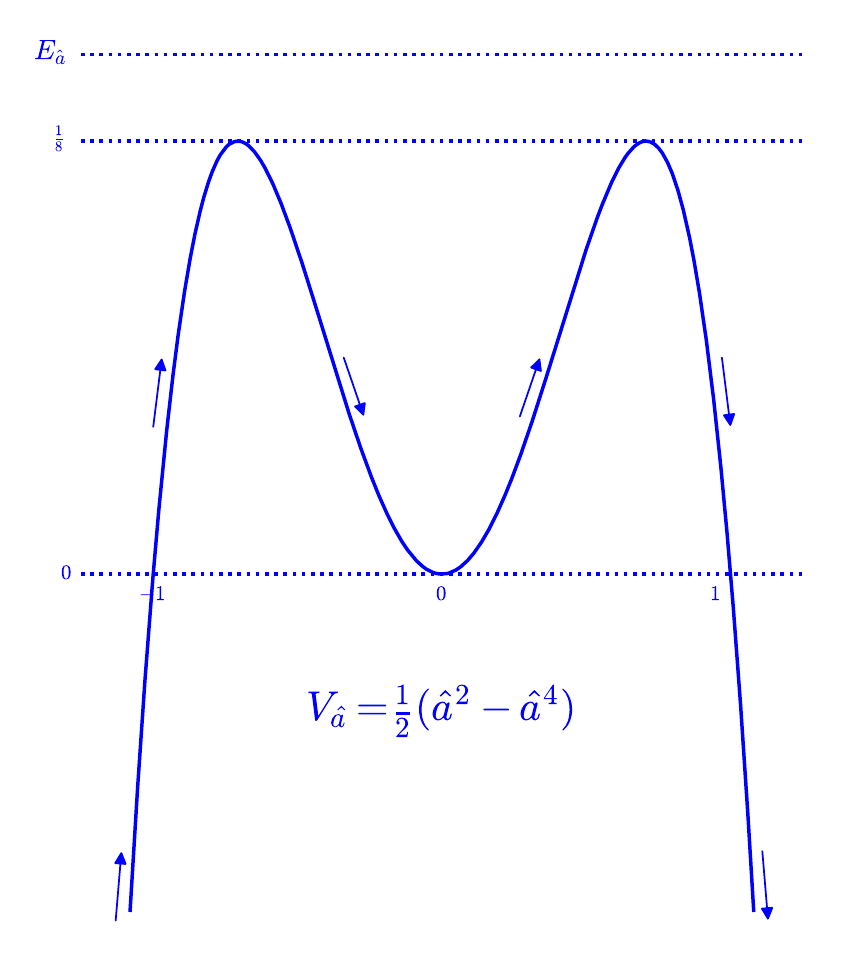}
\includegraphics[scale=0.5]{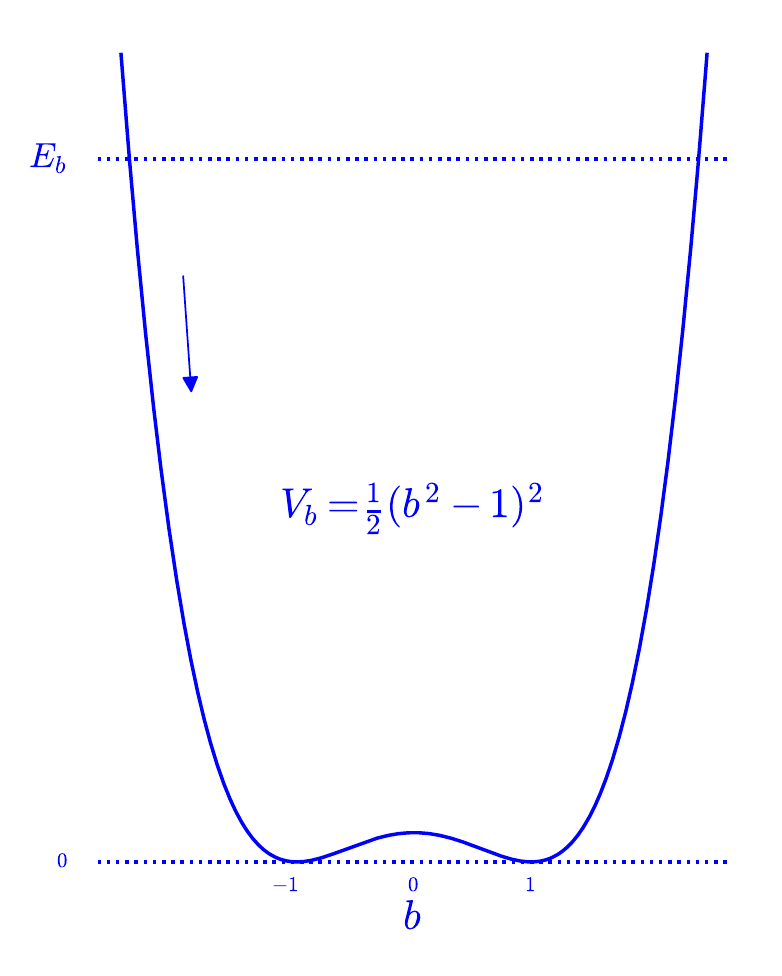}
\caption{The potentials.
$E_{\hat a}$ and $E_{\hat b}$ are not to scale.
\label{fig:potentials}}
\end{figure}

Tha action (\ref{eq:SMGR}) of the $\Spin(4)$-symmetric fields is
\eq
\label{eq:totalS1}
\begin{split}
\frac1{\hbar} S = 2\pi^{2}\int 
&
\left(
\frac{3}{\kappa\hbar} 
\left[
- (\partial_{T}a)^{2}
+a^{2}
\right]
-\frac{1}{8} \lambdaH^{2} \vH ^{4} a^{4}
\right .
\\
&
\left .
{}+
\frac{3 }{2\gtwo ^{2}}
\left [
\left(\partial_{T}b\right)^{2}-
\left(b^{2}-1\right)^{2}
\right]
\right)
\,dT
\end{split}
\en
Define a time scale $\tI$, a dimensionless scale factor $\hat a(T)$ and 
a dimensionless parameter $\epsI$ by
\eq
\label{eq:tIdef}
\begin{gathered}
\tI = \frac{\sqrt{24}}{(\kappa \hbar)^{1/2} \lambdaH \vH ^{2}}
\qquad
a(T) = \tI \,\hat a(T) 
\qquad
\epsI^{4} = \frac{\kappa\hbar }{2\gtwo ^{2}\tI^{2} }
\end{gathered}
\en
to write the action in terms of dimensionless quantities
\eq
\label{eq:actionahatb}
\begin{gathered}
\begin{aligned}
\frac1{\hbar} S
=
\frac{6\pi^{2}}{\gtwo ^{2}}
\int 
\bigg(
-&
\epsI^{-4}
\left[
\frac12
 (\partial_{T}\hat a)^{2}
- V_{\hat a}
\right]
\\
&   {}+
\frac{1}{2}
\left(\partial_{T}b\right)^{2}-V_{b}
\bigg)
\,dT
\end{aligned}
\\[1ex]
V_{\hat a}= \frac12(\hat a^{2} - \hat a^{4})
\qquad
V_{b}= \frac12 (b^{2}-1)^{2}
\end{gathered}
\en
This is a pair of anharmonic oscillators,
$\hat a$ in an inverted potential.
The potentials are graphed in Figure~\ref{fig:potentials}.

%\section{Classical equations of motion}
The classical equations of motion of the action (\ref{eq:SMGR}) 
become the anharmonic oscillator equations
\eq
\label{eq:eqmotion1}
\partial_{T}^{2} \hat a +\hat a -2\hat a^{3} =0
\qquad
\partial_{T}^{2} b -2 b +2b^{3} =0
\en
subject to the Wheeler-Dewitt constraint
\eq
\label{eq:eqmotion2}
\begin{gathered}
H =   - \epsI^{-4} E_{\hat a} +E_{b}= 0
\\
E_{\hat a} =\frac12 (\partial_{T}\hat a)^{2} + V_{\hat a}
\qquad
E_{b} = \frac12 (\partial_{T}b)^{2} +V_{b}
\end{gathered}
\en
$E_{\hat a}$ and $E_{b}$ are the conserved energies.
$H$ generates translation in time.
The $H=0$ constraint expresses reparametrization invariance.

%\section{Numbers}
%\label{sect:numbers}

The  coupling constants $\gtwo $, $\lambdaH$, and $\vH $ are 
fixed by
the Fermi constant $G_{\mathrm{F}}$ and the $W$ and Higgs masses \cite{PDG}
\eq
\begin{gathered}
\begin{alignedat}{2}
G_{\mathrm{F}} &= \frac1{\hbar^{2}\sqrt2\, \vH^{2}} &&= 1.1663787(6)\ntimes 10^{-5}\, \GeV^{-2}
\\
m_{W} &= \hbar \frac{\gtwo \vH}{2} &&= 80.379(12) \,\GeV
\\
m_{H} &= \hbar \lambdaH \vH &&= 125.10 (14) \,\GeV
\end{alignedat}
\\[1ex]
\hbar \vH ={246\,\mathrm{GeV}}
\qquad
\gtwo  = 0.653
\qquad
\lambdaH = 0.508
\end{gathered}
\en
Let
$\tP =  (\hbar\kappa )^{\frac12} = 2.70 \ntimes 10^{-43}\,\mathrm{s}$
be the gravitational time scale
and $\tW = {\hbar}/{m_{W}}  = 8.19 \ntimes 10^{-27}\,\mathrm{s}$
the weak time scale.
Then $\tI$ defined in (\ref{eq:tIdef}) is the seesaw scale
\eq
\tI  =1.03 \, \frac{\tW^{2}}{\tP} 
= 2.55 \ntimes  10^{-10}\,\mathrm{s} 
\en
(which is $7.64\,\mathrm{cm}$, baseball-scale).
$\epsI$ is the seesaw ratio
\eq
\epsI
= 1.03 \, \frac{\tP}{\tW}
= 1.05 \, \frac{\tW}{\tI} = 3.39 \ntimes 10^{-17}
\en
The units of action for $\hat a$ and $b$ in (\ref{eq:actionahatb}) are
\eq
\frac{\hbar_{\hat a}}\hbar = \frac{6\pi^{2}}{\gtwo ^{2}} \epsI^{-4}  =139\, \epsI^{-4}
\qquad
\frac{\hbar_{b}}\hbar = \frac{6\pi^{2}}{\gtwo ^{2}} =139
\en
so classical physics is accurate if
$E_{b}= \epsI^{-4}E_{\hat a}\gg 1$.

%\section{Stability of $\phi = 0$}

$\Spin(4)$ symmetry implies $\phi = 0$
which, in the absence of a gauge field, is 
the unstable equilibrium of the Higgs potential. 
The $\Spin(4)$-symmetric solution 
can be physical only as long as $\phi =0$ is stable.
The scalar field action (\ref{eq:SMGR}) expanded to quadratic order 
in $\phi(x)$ is
\eq
\begin{gathered}
\begin{aligned}
\int
\big[
-
&
\left(D_{\mu}\phi\right)^{\dagger} 
\left(D^{\mu}\phi\right)
+\frac12 \lambdaH^{2} \vH ^{2} \phi^{\dagger}\phi 
\big]\, a^{4} \dVol dT
\\
&=
\int \big[\left(\partial_{T}\phi\right)^{\dagger}\left( \partial_{T}\phi\right)
-\cV(\phi)
\big]\, a^{2} \dVol dT
\end{aligned}
\\
\cV(\phi) =  \hat g^{jk}\left(D_{j}\phi\right)^{\dagger}\left( D_{k}\phi\right)
-\frac12 a^{2}\lambdaH^{2} \vH ^{2} \phi^{\dagger}\phi
\end{gathered}
\en
$\hat g_{jk}$ being the metric of the unit 3-sphere
and $\dVol$ its volume element.
The stability condition is 
\eq
\label{eq:stabilitycondition}
0 \le \int_{S^{3}} \cV(\phi)\, \dVol
\en
for all perturbations $\phi(x)$.
First consider the zero-mode perturbations,
$\phi$ constant in space.
Then
\eq
\begin{gathered}
D_{k}\phi = (\partial_{k}+B_{k})\phi= (1+b) \gamma_{k}\phi
\\
\hat g^{jk}\left(D_{j}\phi\right)^{\dagger} \left(D_{k}\phi \right)
= \frac34(1+b)^{2}\phi^{\dagger}\phi
\end{gathered}
\en
so the zero-mode stability condition is
\eq
\label{eq:weakstability}
0\le \frac34(1+b)^{2} -  \frac12 a^{2}\lambdaH^{2}\vH ^{2} 
\en
$b(T)$ oscillates in the  
potential $V_{b}$ between $\pm b_{\max}$ where
\eq
V_{b}(b_{max}) = \frac12 (b_{\max}^{2}-1)^{2} =E_{b}
\en
Stability of $\phi=0$ at time scales longer than the period of 
$b$ oscillation is the time average of condition (\ref{eq:weakstability}).
\eq
\label{eq:stability2}
0\le \frac34 +\frac34 \langle b^{2}\rangle - \frac12  a^{2}\lambdaH^{2}\vH ^{2} 
\en
This is satisfied when $a(T)$ is small.
When $a(T) $ grows large enough that
condition (\ref{eq:stability2}) is violated,
the 
$\phi=0$ solution becomes unstable.
The Higgs field then starts down from the unstable equilibrium at 
$\phi=0$ towards its 
vacuum expectation value.
This is the beginning of the electroweak transition,
at scale factor $a(T)=a_{\EW}$  where equality holds in 
condition (\ref{eq:stability2}).

For an extremely rough estimate of the actual value of $a_{\EW}$
suppose that the electroweak transition began at redshift ${\approx} 10^{10}$
to within some orders of magnitude.
Suppose that the present scale factor $a_{0}$ is the Hubble time $\tH=  4.55 \ntimes 
10^{17}\,\mathrm{s}$ again to within some orders of magnitude.
Then  $a_{\EW}$ is ${\approx} 10^{7}\,\mathrm{s}$.
Combine this in (\ref{eq:stability2}) with 
$\lambdaH\vH = m_{\Higgs}/\hbar \approx (10^{-27}\,\mathrm{s})^{-1}$
to get $\langle b^{2}\rangle \approx 10^{68}$.
So the solutions that might be physically interesting will have 
$E_{b}$ quite a large number, something like $(10^{68})^{2}$
and $a_{\EW}$ given by equality in (\ref{eq:stability2}) is 
\eq
\label{eq:stability3}
 a_{\EW}  = 
\left(\frac32  \frac{\langle b^{2}\rangle}{\lambdaH^{2}\vH 
^{2}}\right)^{1/2}
=
\left( \frac{3\langle b^{2}\rangle}2 \right)^{1/2}
\frac{\hbar}{m_{H}}
\en

A complete analysis of the stability condition (\ref{eq:stabilitycondition})
for all modes of the perturbation $\phi (x)$
is carried out in \cite{Note1}.
The result is that the zero-mode is the first mode to become unstable 
as $a(T)$ increases.
So equation (\ref{eq:stability3}) remains valid.

%\section{Solution of the $b$ oscillator by an elliptic function}

The $b$ oscillator is solved by integrating the energy 
equation (\ref{eq:eqmotion2}).
Change variables from $T$, $b(T)$ to $\zb$, $\yb(\zb)$
\eq
T = \epsilonb  \zb
\qquad
b(T)= b_{\max}\yb(\zb)
\qquad
\epsilonb = (8E_{b})^{-1/4}
\en
$\yb$ oscillates between 
$\pm 1$.
The energy equation (\ref{eq:eqmotion2}) is
\eq
\label{eq:cneqa}
\left(\frac{d\yb}{d\zb}\right)^{2} = (1 -\yb^{2})(1-\kb^{2}+\kb^{2}\yb^{2})
\qquad
\kb^{2} = \frac12 +\epsilonb^{2}
\en
solved by the Jacobi elliptic function
$y(z)=\cn(z,k)$ \cite{gradshteyn1996table,DLMF}
%\cite[Chapter 22 and Section 19.2]{DLMF}
%\cite[sections 8.11,14,16]{gradshteyn1996table}
\eq
z = \int_{\cn(z,k)}^{1} \frac{dy}{\sqrt{ (1 -y{}^{2})(1-k^{2}+k^{2}y{}^{2})}}
\en
$\cn(z,k)$ is doubly periodic in the complex variable $z$
with real period $4K$ and imaginary period $4iK'$
\eq
\begin{gathered}
\begin{array}{c|@{\;\;\;}c@{\;\;\;}c@{\;\;\;}c@{\;\;\;}c@{\;\;\;}c}
z        &-2K   & -K  & 0   & K   & 2K \\
\hline
\cn(z,k) & -1 & 0 & 1 & 0 & -1
\end{array}
\\[1ex]
\cn(z + 2K,k) = \cn(z+2i K',k) = - \cn(z,k)
\end{gathered}
\en
$K$, $K'$ are the complete elliptic integrals of the 
first kind
\eq
K(k) = \int_{0}^{1} \frac{dy}{\sqrt{ (1 -y{}^{2})(1-k^{2}+k^{2}y{}^{2})}}
\en
$K'(k) = K(k')$, $k^{2}+k'{}^{2} =1$.
Since $\epsilonb$ is ${\approx} 10^{-34}$,
\eq
\begin{gathered}
\kb = 1/{\sqrt2}
\qquad
b(T) = (2 E_{b})^{1/4}  \cn(\epsilonb^{-1} T, \kb)
\\[1ex]
b(T+ 4 \epsilonb \Kb) = b(T+ 4 i \epsilonb \Kb)  = b(T)
\\[1ex]
\Kb = \Kb' = K(1/\sqrt2) =\frac{\Gamma(1/4)^{2}}{4 \pi^{1/2}} = 
1.854075\ldots
\end{gathered}
\en
The time average of $\cn^{2}(z,k)$ over a real cycle gives
$\langle b^{2}\rangle$
\eq
\langle \cn^{2}\rangle = \frac{\pi}{2K^{2}}
\qquad
\langle b^{2}\rangle = (2 E_{b})^{1/2} \frac{\pi}{2K^{2}}
\en
%
%\section{Cosmological temperature}
%
Equation (\ref{eq:stability3}) now becomes
\eq
\label{eq:aEW}
 a_{\EW} 
=  \frac{(3\pi )^{1/2} (2E_{b})^{1/4}}{2K}
\frac{\hbar}{m_{H}}
=   \frac{(6\pi)^{1/2} }{4K\epsilonb} 
\frac{\hbar}{m_{H}}
\en
$E_{b}$ determines $a_{\EW}$ and vice versa,
so $a_{\EW}$ can be taken as the free parameter of the solution.

The cosmological gauge field $b(T)$ is periodic 
in imaginary proper time with period $T \sim T + 4i \Kb \epsilonb $ .
In co-moving time $t$,
given by $dt= a(T)dT$ so that the space-time metric is
$g_{\mu\nu}dx^{\mu}dx^{\nu} = -dt^{2} + a^{2} ds^{2}_{S^{3}}$,
the imaginary time period is $t \sim t+ 4i \Kb  \epsilonb a$.
The cosmological gauge field therefore acts as a thermal bath.
The period $4\Kb \epsilonb  a $ in imaginary co-moving time is
the inverse temperature,
\eq
\frac{\hbar}{\kB T_{\SU(2)}} = 4\Kb \epsilonb  a 
\en
Using equation (\ref{eq:aEW}) for $a_{\EW}$, the temperature is
\eq
\kB T_{\SU(2)}
= \frac{m_{H}}{(6\pi)^{1/2}} \frac{a_{\EW}}{a}
= 28.8\,\GeV \;  \frac{a_{\EW}}{a}
\en
At $a=a_{\EW}$, the onset of the electroweak 
transition, the temperature is
\eq
\begin{gathered}
\kB T_{\EW} =  28.8\,\GeV
\qquad
T_{\EW} = 3.34 \ntimes 10^{14} \,\mathrm{K}
\end{gathered}
\en
independent of the value of $a_{\EW}$.

%\section{Solution of the $\hat a$ oscillator}

To solve the $\hat a$ oscillator, let
\eq
\hat a = (2E_{\hat a})^{1/4} u
\quad
\epsilona = (2 E_{\hat a})^{-1/4}
\quad
\ka^{2}= \frac12 + \frac14 \epsilona^{2}
\en
The interesting values of $E_{\hat a}^{1/4}= \epsI
E_{b}^{1/4}$ are numbers ${\approx}10^{17}$, so
$\ka = 1/\sqrt2$.
In dimensionless co-moving time $\hat t=t/\tI$, $d\hat t = \hat a dT$, the energy equation 
(\ref{eq:eqmotion2}) is
\eq
\left(u \frac{du}{d\hat t}  \right)^{2} +\epsilona^{2}u^{2} -u^{4}  = 1
\en
The solution is
\eq
\label{eq:aoft}
\begin{gathered}
u = \sqrt{\left (e^{2{\hat t}}-1\right)\left(1-\ka^{2}  +\ka^{2} e^{-2{\hat t}}\right)}
=\sqrt{\sinh\left(2\hat t\right)}
\end{gathered}
\en
with the relation to conformal time $T$ given by
\eq
\label{eq:Tt}
e^{-\hat t} = \cn(\epsilona^{-1} T,\ka)
\en
The scale 
factor $\hat a$ ranges from $0$ to $\infty$
as co-moving time $\hat t$ ranges from $0$ to $\infty$
and conformal time $T$ ranges from $0$ to $\epsilona \Ka$
where $\Ka = K(\ka) = K= 1.854075\ldots$.
The time scale of the $b$ oscillations
is much shorter than the expansion time scale, $\epsilonb/\epsilona =\epsI/\sqrt2 = 2.40\ntimes 10^{-17}$.

Combining (\ref{eq:aEW}), (\ref{eq:aoft}), and (\ref{eq:Tt}) gives
\eq
\begin{alignedat}{2}
a_{\EW} &=     0.504\,(2E_{\hat a})^{1/4}\tI
\qquad&
\hat t_{\EW} &= 0.126
\\
t_{\EW} &= 3.21\ntimes 10^{-11}\,\mathrm{s}
\qquad&
T_{\EW} &= 0.501\,\epsilona
\end{alignedat}
\en
(Here $T_{\EW}$ is proper time at $a=a_{\EW}$, 
not temperature.)

%\section{Discussion}

%\subsection{Electroweak transition}

The next step
will be to find a classical solution for cosmology after the onset of the
electoweak transition, for $a>a_{\EW}$.
The solution will include the $\U(1)$ gauge field along with the $\SU(2)$ gauge field.
The gauge group will be $\U(2)$.
In $\U(2)_{L}\times \U(2)_{R}$
the subgroup consisting of the $(g_{L},g_{R})$ with 
$\det g_{R} = \det g_{L}$ acts on $S^{3}=\SU(2)$  by
$g_{\hat x} \mapsto g_{L} g_{\hat x} g_{R}^{-1}$.  
The symmetry group of the nonzero 
Higgs field $\begin{psmallmatrix} \phi^{1} \\ 0\end{psmallmatrix}$ 
is the $\U(2)$ subgroup
$
g_{L} = 
\begin{psmallmatrix}
1 & 0\\
0 & \det g_{R}
\end{psmallmatrix}
$.
There should be a classical $\U(2)$-symmetric solution describing the 
descent of the Higgs field to its vacuum expectation 
value.

Questions that will arise include: how much expansion will take place 
during the descent?  how much spatial anisotropy will be introduced?
will the cosmological gauge field violate CP symmetry during the 
descent?

The temperature $T_{\EW} = 3.34 \ntimes 10^{14} \,\mathrm{K}$
suggests that $a_{\EW}$ is at redshift ${\approx} 10^{14}$.  
Considerable expansion will be needed during the electroweak 
transition.  The solution (\ref{eq:aoft}) for $a(t)$ has an 
inflationary epoch $a \sim e^{t/\tI}$ but
it does not set in until well after the 
$\Spin(4)$-symmetric solution becomes 
unstable at $\hat t_{\EW}  = 0.126$.
The $\U(2)$-symmetric solution will need to 
spend some time in the inflationary neighborhood.

The $\U(2)$-symmetric solution will be spatially homogeneous because
$\U(2)$ takes every point in $S^{3}$ to every other.
But the little group at a point in space is a $\U(1)$
that leaves a preferred direction invariant.
The $\U(2)$-symmetric solution will be anisotropic.
The question is how much anisotropy remains at large time.

The cosmological $\SU(2)$ gauge field $b(T)$ is $P$-odd and $C$-even
so it is $CP$-odd.
In the $\Spin(4)$-symmetric solution 
the oscillations are even in $b \rightarrow -b$.
If the $b$ oscillations are critically damped or over-damped
in the $\U(2)$ solution, then
the very last
stage of the descent to 0 will break the $b \rightarrow -b$
symmetry and thus break CP.

%\subsection{Thermal quantum corrections}

A second pressing task is to evaluate the corrections to the
$\Spin(4)$-symmetric action due to thermal quantum fluctuations
of all the Standard Model fields
to check whether the classical $\Spin(4)$-symmetric solution is materially affected.
The temperature $T_{\SU(2)}$ is inversely proportional to 
$a(t)$ so there is a question
how far back before $a=a_{\EW}$ can the solution obtain before 
corrections from unknown high energy physics become significant.

%\subsection{Basic questions}

Eventually there will be the question of the origin of the 
$\Spin(4)$ symmetry.  Why should the $\SU(2)$ gauge bundle
be isomorphic to 
the spin bundle at early time?  And why should $E_{b}$ be such a large 
number?

%\subsection{My agenda}

The $\Spin(4)$-symmetric classical cosmology is part of a project to 
find a classical solution of the Standard Model combined with General 
Relativity that is analytic in time and that captures the
observed macroscopic features of cosmology.
The project is an attempt at a top-down,
first-principles calculation of a complete macroscopic cosmology
within the Standard Model combined with General Relativity.
The hope is that the macroscopic classical cosmology
can be refined by microscopic corrections
into a quantitatively testable theory.

The first part of the project \cite{ friedan2019cosmology} was an $\SO(4)$-symmetric classical solution 
that captured some qualitative features of late-time cosmology:
a big bang followed by decelerating expansion followed by
accelerating expansion, driven by purely gravitational dark matter and
energy .
The hope is that the $\U(2)$-symmetric solution will interpolate between the 
$\Spin(4)$-symmetric cosmology at early time and the 
$\SO(4)$-symmetric cosmology at late time.

My particular agenda in this work
is to find evidence in support of a speculative fundamental theory of 
physics in which the laws of physics are produced by the
2d renormalization group \cite{ friedan2019cosmology}.
In the pure gravity simplification of this theory,
the space-time metric $g_{\mu\nu}$ is the coupling of a 2d quantum 
field theory,
and the 2d rg drives $g_{\mu\nu}$ to a 
solution of $R_{\mu\nu}=\nabla_{\mu}v_{\nu}+\nabla_{\nu}v_{\mu}$.
This is an rg fixed point because the metric changes by $R_{\mu\nu}$ 
which is just an infinitesimal reparametrization of space-time.
The 2d rg fixed point equation is General Relativity
extended by the source term $\nabla_{\mu}v_{\nu}+\nabla_{\nu}v_{\mu}$.
The source term provides dark matter and energy \cite{ friedan2019cosmology}.
In the full theory, the classical equations of motion of the 
Standard Model combined with General Relativity are extended by source terms 
corresponding to all the gauge symmetries of the theory.
The present work started out by
investigating the $\Spin(4)$-symmetric equations of 
motion with such source terms,
but the source terms turned out to be identically zero in the solution.
So the $\Spin(4)$-symmetric solution 
is presented here without connection to the 2d rg program
of \cite{ friedan2019cosmology}.
Later,
if a comprehensive macroscopic cosmology with the 2d rg source terms
can be derived successfully,
then global properties of the 2d rg flow 
might be looked to for
explanation of early-time $\Spin(4)$-symmetry
and the magnitude of the gauge field oscillations.

\begin{acknowledgments}
This work was supported by the Rutgers New High Energy Theory Center
and by the generosity of B. Weeks.
I am grateful
to the  Mathematics Division of the 
Science Institute of the University of Iceland
for its hospitality.
\end{acknowledgments}

%\vspace*{4ex}

\raggedright
\bibliography{Literature}

\end{document}